\documentclass[twocolumn,aps,showpacs,prbprintnumbers,fleqn,superscriptaddress]{revtex4}
\usepackage{epsfig}
\usepackage{graphicx}
\usepackage{amsfonts}
\usepackage{subfigure}
\usepackage[dvips]{color}
\newcommand{\ig}{\includegraphics}
\newcommand{\ct}{\cite}
\newcommand{\bi}{\bibitem}

\newcommand{\nd}{\noindent}

\newcommand{\ket}{\rangle}
\newcommand{\non}{\nonumber}

\newcommand{\be}{\begin{equation}}
\newcommand{\ee}{\end{equation}}
\newcommand{\ba}{\begin{eqnarray}}
\newcommand{\ea}{\end{eqnarray}}

\begin{document}

\title{Reverse quenching in a one-dimensional Kitaev model}
\author{Uma Divakaran}
\email{udiva@iitk.ac.in}
\author{Amit Dutta}
\email{dutta@iitk.ac.in}
\affiliation{Department of Physics, Indian Institute of Technology, Kanpur 
208 016, India}

\date{\today}
\begin{abstract}
We present an exact result for the non-adiabatic transition probability and
hence the defect density in the final state of a one-dimensional Kitaev model
following a slow quench of  the  parameter $J_-$, which estimates the anisotropy
between the interactions, as  $J_-(t)\sim -|t/\tau|$.
Here, time $t$ goes from $-\infty$ to $+\infty$ and $\tau$ defines the rate 
of change of the Hamiltonian.
In other words, the spin chain initially prepared in its ground state 
is driven by changing $J_-$ linearly in time up to the 
quantum critical point, which in the model considered here occurs at at $t=0$, 
reversed and then gradually decreased 
to its initial value at the same rate. 
We have thoroughly compared the reverse quenching with its counterpart
forward quenching $i.e.,$ $J_-\sim t/\tau$.
Our exact calculation
shows that the probability of excitations is zero for the wave vector at which 
the  instantaneous energy gap is zero at the critical point $J_{-}=0$
as opposed to the maximum value of unity in the forward quenching.
It is also shown that the defect density 
in the final state following a reverse quenching,  we propose here, 
is nearly half of the defects generated in
the forward quenching. 
We argue that the defects produced when the system reaches the 
quantum critical point gets redistributed in the wave vector space
at the final time in case of reverse quenching whereas it keeps on increasing 
till the final time in the forward quenching. We  study the entropy 
density
and  also the time evolution of the diagonal entropy density in the case of the 
reverse quenching and compare it with the forward case. 
\end{abstract}

\maketitle
\section{Introduction}
Studies of quantum  phase transitions in quantum many particle systems 
have always been a fascinating area of research in condensed matter
physics \ct{sachdev99,dutta96}.
While a plethora of theoretical works have been performed on statics of 
quantum phase transitions (QPT), the dynamics of a quantum system passing through a quantum
critical point has caught the 
attention of researchers only recently \ct{kadowaki,sengupta04,zurek05,damski06,polkovnikov05,polkovnikov07,cucchietti07, levitov06,mukherjee07,divakaran07,
sengupta08,sen08,dziarmaga06,dziarmaga05,santoro08,patane08,mukherjee08,
divakaran08,caneva08,viola08,dziarmaga08,divakaran09,mukherjee09,osterloh07,
bermudez08}.
Understanding quantum dynamics
happens to be a challenging problem  as the physics of 
equilibrium quantum phase transitions gets coupled to the
non-equilibrium dynamics of correlated systems. Theoretical techniques
used in studying the above dynamics are often borrowed from similar studies in
quantum optics, $e.g$., the Landau-Zener transition \ct{landau,sei}.

The dynamical evolution can be initiated in a quantum system  either by a 
sudden change  of a parameter in the 
Hamiltonian which is called a sudden quench \ct{sengupta04}, or by 
a slow quenching of a parameter\ct{zurek05,polkovnikov05}. The effect of
the passage through  a quantum critical point is manifested in the 
eventual dynamics of the system. 
The relaxation time ($\xi_t$) of the system, defined
as the time taken by the system to come back to its equilibrium state after a
small perturbation,
diverges at the critical point as $\xi_t \sim \delta^{-\nu z}$ where $\delta$ measures
the deviation from the quantum critical point and $\nu$ and $z$ are the 
corresponding correlation length and dynamical critical exponents, respectively.
The divergence of the relaxation time is an artifact of vanishing energy
gap $\Delta$ ($\sim \xi_t^{-1}$) between the ground and the first excited 
state of the Hamiltonian 
near a quantum critical point.
This divergence of the relaxation time
forces the system to be infinitely sluggish near the critical
point so that it takes
effectively an infinite time to respond to any change in the external 
parameters thereby causing excitations. 
The recent discovery of ultra cold atoms which has facilitated the
experimental implementation of various Hamiltonian models 
\ct{bloch07,duan03} and thus 
the verification of results of quantum dynamics,
has enormously accelerated the theoretical research  in this field.
Here, we are interested in a slow and linear variation of the
quenching parameter and estimate  various quantities such as density 
of defects and the local entropy density 
in the final state of the system following a quench, 
as a function of the quenching rate $\tau$. 

It is well known that for a $d$-dimensional
system which is initially prepared in its ground state and is  
quenched through a quantum critical point by linearly varying a parameter as $t/\tau$, 
the density of defects ($n$)
satisfies the Kibble-Zurek(KZ) scaling \ct{kibble76,zurek96,zurek05,polkovnikov05,polkovnikov07} given
by $n \sim \tau^{-d \nu /(\nu z+1)}$ where $\nu$ and $z$ are the critical exponents defined above. 
The Kibble-Zurek scaling has been verified in various exactly
solvable spin models \ct{zurek05,levitov06,mukherjee07,divakaran07,dziarmaga05} and
in a system of interacting bosons undergoing superfluid to insulator transitions \ct{polkovnikov05}.
The above KZ scaling relations gets modified when the system is quenched through a multicritical point
\ct{divakaran09}, across a gapless phase \ct{sengupta08,santoro08}, along a gapless line \ct{divakaran08,
viola08} or for quenching with a non-linear rate \ct{sen08}.
Studies on quenching dynamics have also been generalized to quantum spin chains with quenched
disorder \ct{ dziarmaga06}, 
to systems in presence of white noise\ct{osterloh07}, 
to systems with  infinite range interactions \ct{caneva08},
to an open system coupled to a heat bath \ct{patane08}, 
and also to quantum spin chains
driven by an oscillatory magnetic field \ct{mukherjee09} .
The effect of edge states on the defect production has also been studied
\ct{bermudez08}.
It is worth mentioning
here that the defect production has been studied experimentally for a 
rapid quench in a spin-1 bose condensate \ct{sadler06}.

In this paper, we study the effect of the reversal of the
quenching path right 
at the quantum critical point on the density of defects. This
is achieved 
by increasing the  quenching parameter from time $-\infty$ to its value
at the quantum critical point and the
bringing it back at the same rate to its initial
value at the final time $i.e.,~t \to \infty$.
We call this quenching scheme
as reverse quenching whereas the other scheme in which the quenching parameter
is monotonically increased from 
time $-\infty$ to $+\infty$ through a quantum critical point will
be referred to as the forward quenching scheme.
 We would also occasionally use the term half quenching for the 
case when the quenching is stopped at the quantum critical point.

At the outset, let us  discuss the motivation behind our study.
The forward quenching scheme has been applied extensively for the entire range in time from $t =-\infty$ to $\infty$ and defects generated in the final
state has been estimated. In the present work, we drive the system linearly right up to
the quantum critical point and then let it retrace its path. In our calculation,
we find that the defect generated up to time $t=0$ is approximately half
of the defect generated for a full forward quenching. -
We also seek answer to the question that how the defects 
generated in the
first half of the quenching  get altered under reversal of the path!
We  address  questions like do we have a similar scaling form for the
defect  or how does the magnitude of the
defects in the final state, i.e. at $t \to \infty$  change under reverse quenching?. In
the process, we also provide an exact solution of the Schr\"odinfer equation to find the non-adiabatic transition probability for the reverse quenching.

We also compare our work with that in ref.~\ct{mukherjee08} where
a quantum XY spin 
chain is repeatedly swept through the quantum critical points by varying 
the magnetic field at a linear rate between $-\infty$ and
$\infty$ such that the reversal of path takes place far away from the critical 
points. In the present work, the parameter is increased only upto the quantum
critical point where it is reversed to trace back its path.
To employ the reverse quenching scheme in an appropriate way,
we study the dynamics of a one-dimensional Kitaev model 
where the quantum critical point occurs at $t=0$. Secondly, in ref.~\ct {mukherjee08}, 
the defect density after each 
repetition is estimated using a recursive relation for the non-adiabatic 
transition probabilities 
where the (rapidly varying) cross terms are
ignored because they vanish upon integration over the wave vectors. On the other hand, we present 
here an exact solution of the transition probabilities which include interference terms although the 
results are in fairly good agreement at least qualitatively with the one cycle 
case discussed in ref \ct{mukherjee08}
It is also to be noted that the reverse
quenching dynamics has been studied for a generic two level system in ref.~
\ct{garanin02} and the excitation probability has been calculated within the
framework of perturbation theory. On the other hand, we here generalize the 
quenching scheme to a many particle system
and solve the Schr\"odinger equations exactly.


The paper is organized in the following way: the model and the 
quenching scheme are discussed in section II. 
The main results for the non-adiabatic transition probability, 
defect density and the entropy density are presented in section III. 
We summarize our results in the concluding section whereas the 
calculational details are provided in the appendix.


\section{The model and the quenching Scheme}

Two-dimensional Kitaev model defined on a honeycomb lattice described by the 
Hamiltonian \ct{kitaev06}
\ba
\tilde H=\sum_{n+l={\rm even}}(\sigma_{n,l}^x\sigma_{n+1,l}^x+J_2\sigma_{n-1,l}^y
\sigma_{n,l}^y+J_3\sigma_{n,l}^z\sigma_{n,l+1}^z)
\ea
where $n$ and $l$ defines the column and row indices of the lattice
has been studied extensively due to its exact solvability by
Jordan-Wigner transformation\ct{lieb61}. 
The rich phase diagram of this model has a gapless phase, through which 
the quenching dynamics has been studied recently\ct{sengupta08}.
The one dimensional version of the Kitaev model (with $J_3=0$) given by the 
Hamiltonian
\ct{kitaev06,nussinov} 
\ba
H=\sum_{n=1}^{N}(J_1 \sigma_{2n}^x\sigma_{2n+1}^x + J_2\sigma_{2n-1}^y\sigma_{2n}^y)
\label{kit_ham}
\ea
\nd where $n$ refers to the site index, exhibits a quantum phase transition
at $J_1=J_2$. The above Hamiltonian (\ref{kit_ham}), which is the model
of interest in this work, can be exactly 
diagonalized by standard Jordan Wigner transformation \ct{lieb61}
as defined below
\ba
a_n=(\prod_{j=-\infty}^{2n-1} \sigma_j^z)\sigma_{2n}^y,~~~
b_n=(\prod_{j=-\infty}^{2n} \sigma_j^z)\sigma_{2n+1}^x.~~~
\ea
Here $a_n$ and $b_n$ are independent Majorana fermions at site $n$ \ct{sengupta08}. They 
satisfy the relations like
\ba
a_n^\dag=a_n,~
b_n^\dag=b_n,~
\{a_m,a_n\}=2\delta_{m,n},~\nonumber\\
\{b_m,b_n\}=2\delta_{m,n},~
\{a_m,b_n\}=0.
\ea
Substituting for $\sigma_n^x$ and $\sigma_n^y$ in terms of Majorana fermions
followed by a fourier transformation, Hamiltonian (\ref{kit_ham}) 
can be written as
\ba
H=2i\sum_{k=0}^{\pi}[b_k^\dag a_k(J_1+J_2 e^{ik})+a_k^{\dag}b_k(-J_1-J_2e^{-ik})]\label{kit_map}
\ea
where the fourier component $a_k$, satisfying the standard anticommutation 
relations $\{a_k,a_{k'}^\dag\}=\delta_{k,k'}$ and 
$\{a_k,a_{k'}\}=0$, is defined as
\ba
a_n=\sqrt{\frac{4}{N}}\sum_{k=0}^{\pi}[a_k e^{ikn} + a_k^\dag e^{-ikn}]\nonumber\\
+\sqrt{\frac{2}{N}}[a_0+a_0^\dag+a_{\pi}(-1)^n+a_{\pi}^\dag(-1)^n].
\label{fourierak}
\ea
\noindent The sum over $k$ in Eq.~(\ref{fourierak}) goes only 
for half the Brillouin zone as $a_n's$ are Majorana fermions.
By defining $\psi_k=(a_k,b_k)$, the Hamiltonian (\ref{kit_map}) can then be 
rewritten in a simpler form as
\be
H=\sum_{k=0}^{\pi}\psi_k^\dag H_k \psi_k
\ee
where 
\ba 
H_k=2 i\left[ \begin{array}{cc} 0 & -J_1-J_2e^{-ik} \\
J_1+J_2 e^{ik} & 0 \end{array} \right]. \non \\
& & \label{kit_mat} \ea

The above Hamiltonian can be diagonalized where the eigenvalues are given by
$$\epsilon_k^{\pm}=\pm2\sqrt{J_1^2+J_2^2+2J_1J_2 \cos k}.$$ 
Clearly, the gap vanishes at $J_1=\pm J_2$ for $k=\pi$ and $0$ respectively
with the
critical exponents $\nu$ and $z$ both being equal to unity.
 Feng, Zhang and Xiang \ct{fengprl07} showed that this vanishing
energy gap signals a topological phase transition between the
two phases of the model at $J_1<J_2$ and $J_1>J_2$. Interestingly, this model
can be mapped to a one-dimensional transverse Ising model by a duality
transformation \ct{fengprl07,jhhp77,nussinov08}.

In terms of a new set of basis functions given by

\ba \psi_{1k} &=& \frac{1}{\sqrt{2}} ~\left( \begin{array}{c} 1 \\ i
\end{array} \right) ~~{\rm and}~~ \psi_{2k} ~=~ \frac{1}{\sqrt{2}} ~\left(
\begin{array}{c} 1 \\ -i \end{array} \right), \nonumber \ea
the above Hamiltonian
can be recast to the form 

\ba 
\tilde{H_k}=2 \left[ \begin{array}{cc} \frac {J_{+} + J_{-}}{2} +\frac {J_{+} - J_{-}}{2}\cos k  & - \frac {J_{+} - J_{-}}{2}\sin k \\
-\frac {J_{+} - J_{-}}{2} \sin k & -\frac {J_{+} + J_{-}}{2}- \frac {J_{+} - J_{-}}{2}\cos k \end{array} \right] \label{kit_mat1} 
 \ea
\nd where $J_{\pm}=J_1\pm J_2$. 
We study the dynamics of the spin chain by varying the term  $J_-$ 
of Hamiltonian (\ref{kit_mat1})
using the quenching rule $J_-=-|\frac{t}{\tau}|$
where $t$ varies from $-\infty$ to $+\infty$. 
Here, the quantum critical point occurs at $t=0$ and it is at this point where
the parameter $J_-$ is reversed to bring it back to its initial value at the 
final time.
It is to be noted that the off-diagonal terms in the
Hamiltonian (\ref{kit_mat1}) becomes time-dependent in the present quenching scheme making
the analytical solution difficult. However, the situation can be easily saved by making an
appropriate unitary transformation as shown below: In the limit of large $t$ 
($t \to \pm \infty$), the eigenvectors of the Hamiltonian are 
\ba
|e_{1k}\ket&=&\frac{1}{\sqrt{2(1+\sin(k/2))}}[\cos\frac{k}{2}|\psi_{2k}\ket+(1+\sin\frac{k}{2})|\psi_{1k}\ket]\nonumber\\
|e_{2k}\ket&=&\frac{1}{\sqrt{2(1+\sin(k/2))}}[-\cos\frac{k}{2}|\psi_{1k}\ket+(1+\sin\frac{k}{2})|\psi_{2k}\ket]\nonumber
\ea
where $|e_{1k}\ket$ is the ground state in the limit $t \to -\infty$.
A unitary transformation generated by the matrix $U$ constructed from the
above eigenvectors leads to the final the Hamiltonian suitable for
the present form of quenching and is given by

\ba 
H_k'&=&U^{\dagger} \tilde{H_k} U\nonumber\\
   &=& 2 \left[ \begin{array}{cc} J_{-}(t)\sin(k') &  J_+\cos(k') \\
 J_+ \cos(k') & -J_- (t)\sin(k') \end{array} \right], \non \\
& & \label{kit_mat2} \ea
where the time dependence is now entirely shifted to diagonal terms of the 
Hamiltonian. 
The quantum critical point is at $J_-=0$ for the mode $k'=\pi/2$. Also,
the mode $k'$ in Eq.~(\ref{kit_mat2}) is half of mode $k$ in 
Eq~(\ref{kit_mat}). Henceforth, we will refer $k'$ as $k$ and appropriately
redefine the Brillouin zone. 
 The presence of a single QCP precisely at $t=0$ renders
the analytical calculation easier and so we chose Kitaev model over
other exactly solved models for the present
study. We note that the results of this model can be extended to 
any other Jordan-Wigner solvable models.


\section{Results}
In this section, we shall present the main results of this work. 
The $2\times 2$ reduced  Hamiltonian matrix given in 
Eq.~(\ref{kit_mat2}) can be interpreted as a Landau Zener Hamiltonian \ct{landau,sei} where the diagonal 
elements are the two  bare (diabatic) energy levels which approach each other
and the off-diagonal element $\Delta_k$ is the minimum gap between the 
instantaneous levels of
the Hamiltonian. At time $t=0$, the energy gap
between the instantaneous energy levels vanish for the mode $k=\pi/2$ signaling 
a quantum phase transition mentioned above. 
We shall 
assume that the system is prepared in its 
initial ground state $|e_{1k}\ket$
at $t \to -\infty$. At any instant $t$ during the evolution, a general state 
vector
$|\psi_k(t)\ket$ can be expressed as 
$|\psi_k(t)\ket=c_{1,k}(t)|e_{1k}\ket + c_{2,k}(t)|e_{2k}\ket$ 
where $c_{i,k}(t)$ (i=1,2) denote the time-dependent
probability amplitude for the bare state $|e_{ik}\ket$.

The Schr\"odinger equation describing the evolution of the system is 
\ba
i\frac{\partial}{\partial t}c_{1,k}(t)&=&2J_-\sin(k) c_{1,k}(t) + 2 \cos(k) c_{2,k}(t)
\non\\
i\frac{\partial}{\partial t}c_{2,k}(t)&=&-2J_-\sin(k) c_{2,k}(t) + 2 \cos(k) c_{1,k}(t),
\label{schr1}
\ea
with initial conditions $c_{1,k}(-\infty)=1$, $c_{2,k}(-\infty)=0$ 
and we have set $J_{+}=1$. 
The nonadiabatic transition probability in the final state is
given by $|c_{2,k}(+\infty)|^2$.
The above Schr\"odinger equations are solved exactly 
and the probability of excitation for the $k$-th mode, $p_k$, is 
\ba
p_k&(t \to \infty)=&\frac{1}{4}(1-e^{-2\pi\alpha})\non\\
&\times&\left|\frac{\Gamma(1-i\alpha/2)}{\Gamma(1+i\alpha/2)}+i\frac{\Gamma(1/2-i\alpha/2)}{\Gamma(1/2+i\alpha/2)}\right|^2,
\label{eq_pk}
\ea
where $\alpha=\tau \cos^2(k)/\sin(k)$. 
The density of defects can be obtained by integrating the probability
of excitations $p_k$ over the Brillouin zone and is given by 
\ba
n=\frac{1}{2\pi}\int_{-\pi}^\pi p_k(t \to \infty)~ dk =\frac{1}{\pi}\int_0^\pi p_k(t \to \infty)~ dk .
\ea

The parameter $\alpha$ measures
the effective rate of driving. It is $\alpha$, not $\tau$ which
determines the diabatic $(\alpha \to 0)$ and adiabatic $(\alpha \to \infty)$ 
limit \ct{vitanov99,damski06}. The gap varies with wave vector $k$, so does $\alpha$ and
for the modes close to the critical mode $(k=\pi/2)$, $\alpha \sim k^2 \tau$.
We defer the calculational details to the Appendix.

Let us first  analyze the exact expression  given in Eq.~(\ref{eq_pk}) in 
different limits and compare 
it with $p_k$ obtained
by direct numerical integration of the Schr\"odinger equation (\ref{schr1}).  
It should be noted that the first expression in the modulus squared term of Eq.~(\ref{eq_pk}) is
$$\frac{\Gamma(1-i\alpha/2)}{\Gamma(1+i\alpha/2)}=
\frac{\Gamma(z)}{\Gamma(\bar z)}=\frac{\Gamma(z)}{\overline{\Gamma(z)}}$$
which is a unit vector with argument $-2\theta_1$ where $\theta_1$ is the 
argument of $\Gamma(1+i\alpha/2)$. Similarly the second expression in the 
modulus squared term is also a unit vector with angle $-2\theta_2$ and hence
modulus squared term in Eq.~(\ref{eq_pk}) reduces to
\ba
2+2\sin(2\theta_2-2\theta_1)
\label{gammaterm}
\ea

Therefore, the final expression for the probability of excitations is
\ba
p_k(t\to \infty)&=&\frac{1}{4}(1-e^{-2\pi\alpha})
\times\left|2+2\sin(2\theta_2-2\theta_1)\right|
\label{eq_pk_final}
\ea

\begin{figure}
\ig[height=2.5in]{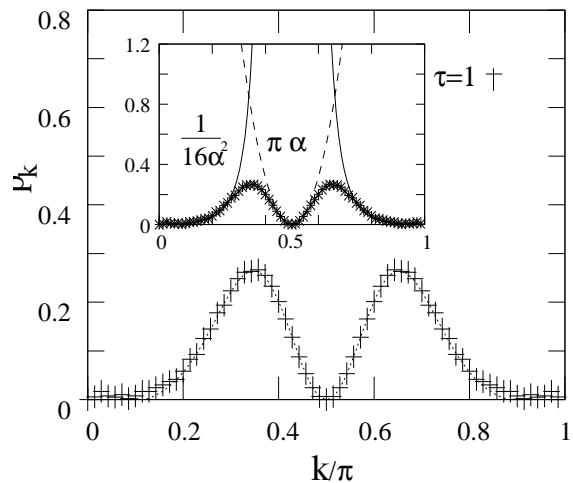}
\caption{Variation of $p_k$ vs k for $\tau=1$. The data with `+' sign  represent the numerical solution whereas the dashed line corresponds to the analytical expression given in Eq.~\ref{eq_pk}.  As explained in the text, the region near $k=\pi/2$ varies linearly with $\alpha=\tau \cos^2 k/\sin k$ whereas that away from $k=\pi/2$
as $1/\alpha^2$. The inset also shows the intersection of the two limits 
 at a particular
mode $k_0$. The dotted line in the inset goes as $\pi \alpha$ and the thin line falls
as $1/16\alpha^2$. }
\label{pkvsk}
\end{figure}
\nd Using the properties of $\Gamma$ function\ct{abram}, we have
\ba
\theta_1=\frac{\alpha}2\psi(1) + \left(\sum_{l=0}^{\infty}\frac{\alpha}
{2(1+l)}-\tan^{-1}(\frac{\alpha}{2(1+l)})\right)\non\\
\theta_2=\frac{\alpha}2\psi(\frac{1}2) + \left(\sum_{l=0}^{\infty}
\frac{\alpha}{2(1/2+l)}-\tan^{-1}(\frac{\alpha}{2(1/2+l)})\right)
\label{eq_theta}
\ea
\noindent where $\psi(z)$ is the well known Digamma function\ct{abram}.
The analytical expression for $p_k$ is obtained by   
substituting  $\theta_1$
and $\theta_2$ in Eq~(\ref{eq_pk_final}).
The result obtained by numerical integration of the Schr\"odinger equation is
presented in Fig.~(\ref{pkvsk}) where we also plot the analytical expression 
after substituting numerically obtained values of $\theta_1$ and $\theta_2$ 
in Eq.~(\ref{eq_pk_final}). 

The behavior of $p_k$ as a function of the wave vector $k$  
can be explained by making resort to the Landau-Zener interpretation discussed before.
For the modes close to $k=0$ 
(which are away from the critical mode $k=\pi/2$), the minimum gap 
$\Delta_k$ is relatively large, or more precisely $\Delta_k^2 \tau >>1$.
Hence, these modes evolve adiabatically remaining close to the 
instantaneous ground state throughout the quenching. 
For the critical mode $k=\pi/2$,
gap is zero or in other words the relaxation time (inverse of the gap)
is diverging which results to the complete freezing of dynamics. 
The system stays in its initial ground state throughout
the evolution which also happens to be the ground state at $t \to +\infty$ 
for the present scheme of quenching. We therefore
encounter a situation where the mode for which instantaneous energy gap is zero 
has simultaneously zero excitation probability
which is in contrast to the forward case where the probability of excitations 
is unity for the critical mode. Similarly,
for the modes near $k=\pi/2$ where the gap is still very small, the system stays closer to
the initial state due to large relaxation time leading to a final state similar to the 
ground state. We therefore conclude that the
transition probability vanishes in either limits $k\to 0$ and $k\to \pi/2$ and 
a peak is expected at a wave vector $k_o$ lying somewhere in the middle as shown in Fig.~(1). 

\begin{figure}
\ig[height=3.2in,angle=-90]{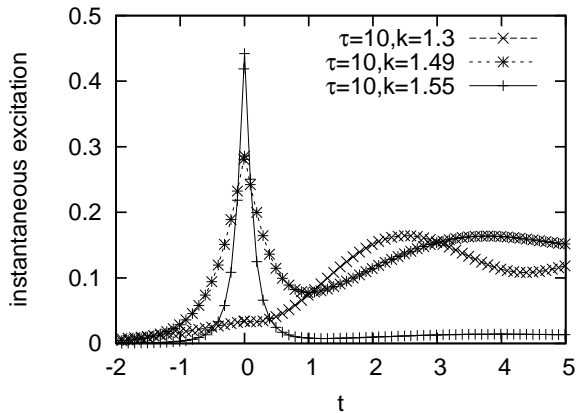}
\caption{Instantaneous excitation probabilities vs time for three different modes. The mode $k=1.49$
is closer to the $\pi/2$ mode and shows a diabatic behavior $i.e.,$
the small gap and large relaxation time makes the system unable to change 
appreciably from the initial state and therefore the instantaneous
excitation decreases in magnitude for $t>0$ when it is retracing its path.
On the other hand, the system tries to follow the instantaneous
ground state  for $k=1.3$ which is in the adiabatic limit and the excitation keeps on
increasing till the effect of finite gap persists. Finally, for the mode $k=1.55$ which
is closest to the critical mode $k=\pi/2$, decrease in instantaneous excitation for 
$t >0$ is prominently shown.}
\label{inst_exctn}
\end{figure}

The instantaneous excitation at an instant $t$ is defined as the probability of finding
the system in the instantaneous excited eigenstate of the Hamiltonian (10) . The variation
of  the instantaneous excitation probability
 as a function of time  presented  in Fig.(\ref{inst_exctn})
reflects the explanation presented above.

In the limit of small $\alpha$ (i.e., either $\tau$ small or $k \to \pi/2$), 
we get a simplified expression
\be
2\theta_2-2\theta_1=-2\alpha\ln2.
\label{smallalpha}
\ee
Substituting Eq.~(\ref{smallalpha}) in Eq~(\ref{eq_pk_final}), we get the expression
for $p_k$ in the small $\alpha$ limit as
\ba
p_k&=&\frac{1}{4}(1-e^{-2\pi\alpha})
\times (2-2\sin(2\alpha\ln2)).
\label{pk_smallalpha}
\ea
which correctly predicts the curve for small $\alpha$ along with the peak of the curve.

It would be useful to calculate $p_k$ in the two extreme limits, namely, $\alpha \to 0$ and 
$\alpha \to \infty$. 

\begin{figure}
\ig[height=3.6in,angle=-90]{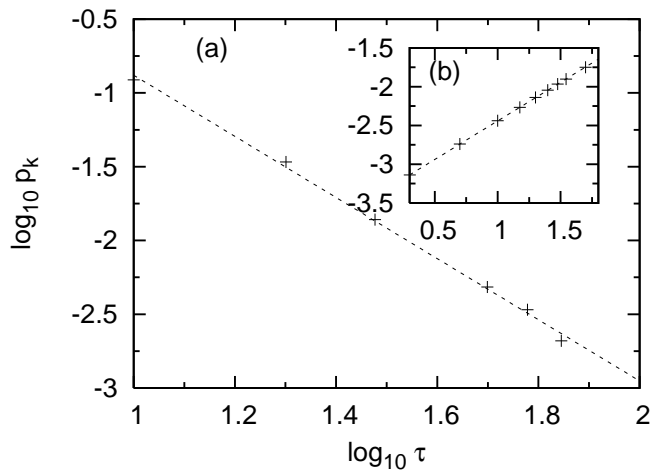} 
\caption{Variation of $\log p_k$ vs $\log \tau$ for two different k values. 
Fig (a) corresponds to k=1.3 where $1/\tau^2$ behavior is expected. 
The dots are the numerically obtained values where as the fitted line 
has a slope -2. Fig (b) (inset), on the other hand is for $k=1.56$ where $p_k$
increase linearly with $\tau$. Once again a log-log plot shows a slope of 
1 as expected from theory.}
\label{differenttau}
\end{figure}

The $\alpha \to 0$ limit can be obtained directly
from Eq.~( \ref{eq_pk}) whereas the large $\alpha$ limit is obtained
from using the asymptotic expansion of the Gamma function in Eq~(\ref{eq_pk}). 
Thus, we have
\ba
p_k&\sim&\pi\alpha ~~~~~~{\rm for}~~\alpha \to 0\non\\
p_k&\sim& \frac{1}{16\alpha^2}~~~~~~{\rm for}~~\alpha \to \infty
\ea
which is also shown in the inset of Fig.~1.
\noindent As discussed already, the $\alpha\to 0$ behavior is expected near the 
critical mode $k=\pi/2$  whereas large $\alpha$ behavior can be observed 
for the modes $k$ away from $\pi/2$. It is also interesting to note that 
the exact solution we present here reduces to the adiabatic transition 
probability scaling $1/\alpha^2$ in the limit of large $\alpha$ 
as expected from the quantum mechanical adiabatic theorem \ct{messiah,sei}.
This feature is more transparent in Fig.~\ref{differenttau} where the 
variation of 
$p_k$ with $\tau$ for two different values of $k$, one near $k=\pi/2$ and the other away
from $k=\pi/2$, is shown depicting the linear increase with $\tau$ and decrease
as $1/\tau^2$ respectively.

\begin{figure}
\ig[height=3.2in,angle=-90]{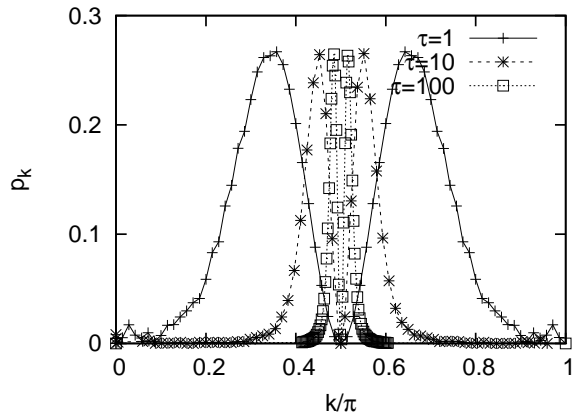}
\caption{The variation of $p_k$ for different values of $\tau$ showing that 
the maximum value of $p_k$ is independent of $\tau$}
\label{varytau}
\end{figure}

The mode $k_0$ at which the peak in $p_k$ occurs is approximately given by the 
point of intersection of two limiting behaviors given in Eq.~15 
 and is given by
$$\frac{\cos^2(k_0)}{\sin(k_0)}=(\frac{1}{16\pi})^{1/3}\frac{1}{\tau}.$$
The value of $\alpha$ at  $k_0$ is given by
$$\alpha(k=k_0)=\sqrt[3]{1/16\pi}$$
which is  independent 
of $\tau$. This implies that the maximum value of $p_k$ is 
independent of $\tau$ as shown in Fig~\ref{varytau}. A rough estimate of this value can be obtained from Eq~(\ref{pk_smallalpha})
after substituting $\alpha$ at $k_0$ and is found to be close to $0.25$.

Let us shift our attention to estimating the density of defects $n$ in 
the final state at $t\to \infty$.
 The variation of density of defects with the rate of quenching $\tau$ is obtained by integrating the probability
of excitations over the Brillouin zone given by Eq.~(13).  We find that $n(\tau)$ as a function of
$\tau$ shows a peak at a particular quenching
rate $\tau_0$ and eventually follows a $1/\sqrt{\tau}$ decay for 
very large $\tau$. 
The $1/\sqrt{\tau}$ behavior is justified by noting the fact
that for very large $\tau$, $k_0$ shifts towards $\pi/2$ where 
$\cos^2(k)/\sin(k)\sim k^2$.
In this large $\tau$-limit therefore, $p_k$ scales as $p_k \sim p_k(\tau k^2)$ 
resulting to the Kibble-Zurek scaling of the defect density given as
$n \sim 1/\sqrt{\tau}$. 
Fig.~5 shows the variation of density of defects with
$\tau$
in the large $\tau$ limit with a $1/\sqrt{\tau}$ behavior whereas the inset of 
Fig.~5 corresponds to the $n$ $vs$ $\tau$ behavior for the entire range of
$\tau$ depicting the peak as described above.

\begin{figure}
\ig[height=3.6in,angle=-90]{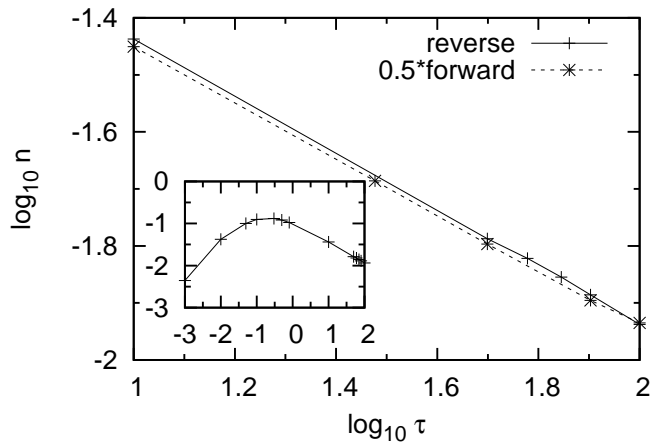}
\caption{The main part of the figure shows the variation of density of defects $n$ for relatively higher values $\tau$ for reverse quenching. 
Also plotted is half times the density of defects produced while forward quenching and 
it is clear that reverse quenching is close to half of the linear quenching. Inset shows
$n$ vs. $\tau$ for a wider range of $\tau$ where a peak is observed for a relatively
smaller value of $\tau$} 
\label{nvstau}
\end{figure}

An interesting observation is that for the present quenching scheme, 
the density of defects in the final state  is close to half of that 
in the forward quenching i.e., the case where $J_-$ is linearly quenched from 
$-\infty$ to $\infty$, see Fig~(\ref{nvstau}). 
This is because the maximum value of $p_k$ in the reverse case
is one-fourth that of the forward case making the area under one of the 
peaks to be close to 
one fourth that of the linear. 

\begin{figure}
\includegraphics[width=0.35\textwidth,angle=-90]{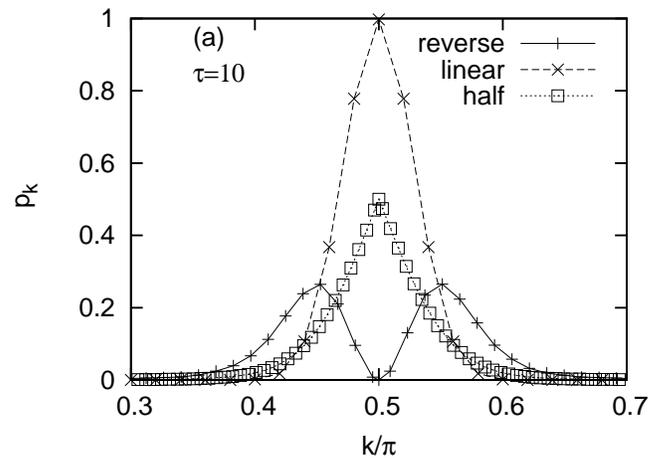}
\includegraphics[width=0.35\textwidth,angle=-90]{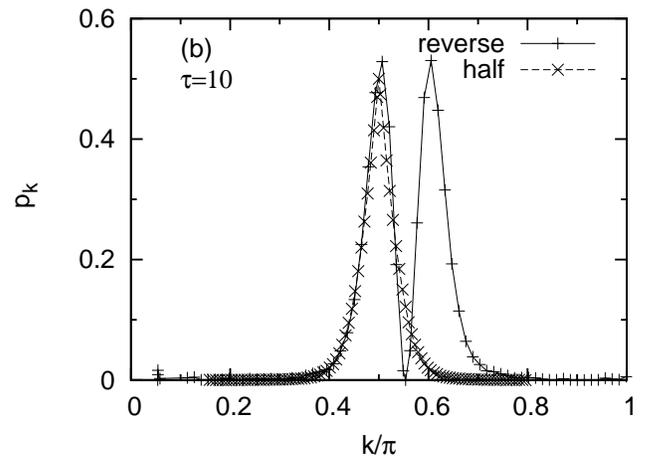}
\caption{a) A comparison between probability of excitations for  
reverse quenching, forward quenching
and $p_k(t=0)$. is presented The range of $k$ is appropriately chosen
for the clarity of presentation.
Fig.~ 6b)shows that the defect generated in the reverse case,  which
is the area under the $p_k$ vs. $k$ curve, is approximately equal to the defect generated at $t=0$. To highlight this, we have doubled $p_k$ for the reverse case and overlapped one of its
peaks with $p_k(t=0$) by appropriate shifting of the $x$-axis.}
\label{all_pkvsk}
\end{figure}

It is also illustrative to compare the non-adiabatic transition probability 
$p_k (t\to +\infty)$ as a function of  $k$ for 
 reverse quenching, forward quenching as well as half quenching i.e., 
$p_k (t=0)$.
Fig.~\ref{all_pkvsk}(a) suggests that the density of defects (area under the
$p_k$ vs $k$ curve) does not change appreciably  
in reverse quenching as compared to the half quenching. However, there is  a reorganization
of $p_k$ in the wave-vector space keeping the density
of defect nearly constant. To justify the above statement, we have doubled
the peaks in the reverse case and appropriately shifted the x-axis
to match one of its peaks with $p_k(t\to 0)$
in Fig \ref{all_pkvsk}(b). The peak is found to match  almost identically to
$p_k(0)$.
In the passing, we note  from Fig.~ (6a),that for very large $\tau$ when only the 
modes close to the critical mode contributes to the defects, 
the density of defects in the forward case is double that of half quenching.

\begin{figure}
\includegraphics[height=3.2in,angle=-90]{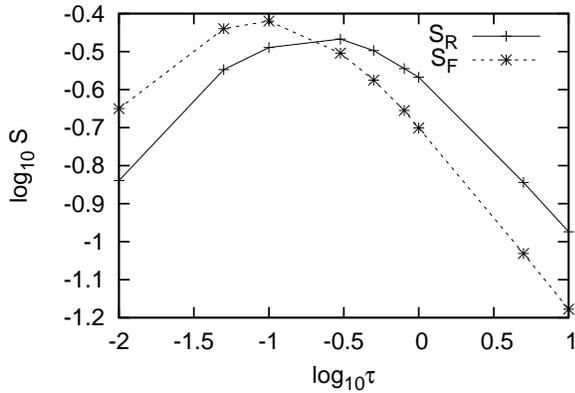}
\caption{Comparison of entropy of reverse and forward quenching.}
\label{entropy}
\end{figure}

We now shift our focus to the  local  von-Neuman entropy density  \ct{levitov06} defined by  

\ba
S=-\frac{1}{\pi}\int_0^{\pi}(p_k \ln p_k+(1-p_k) \ln (1-p_k))~ dk
\label{von}
\ea
generated in the reverse quenching process.
The entropy density is small for small as well as large values of $\tau$
and attains the maximum value at a characteristic time-scale.
A comparison between  the entropy generated in the forward ($S_F)$ and the reverse case ($S_R$) 
is shown in Fig.~\ref{entropy}. 
Although qualitatively the  curves are similar, an interesting
difference is to be pointed out: $S_R$ is less than
$S_F$ in the limit of small $\tau$ whereas $S_R$ exceeds $S_F$ for large $\tau$.
This observation can be explained as follows: the integrand in Eq~(\ref{von})
is maximum when $p_k=0.5$. In reverse case, $p_k$  always remains less than
$0.5$, $i.e.,$ it never reaches the maximally disordered state. 
For the forward case in the small $\tau$ (non-adiabatic) limit,
$p_k$ is close to $0.5$ in a larger region and hence entropy is large. In other words,
the final state is more locally ordered following a reverse quenching than a forward quenching
in the limit $\tau \to 0$.  On the other hand, for the large $\tau$ limit in the forward case, 
$p_k$ increases sharply near the critical mode $k=\pi/2$  and non-negligible only for wave vectors
close to $\pi/2$ resulting to relatively smaller 
entropy density. 

 In a recent work, Barankov and Polkovnikov \ct{polkovnikov082} have proposed the
concept of diagonal entropy given by
\ba
S_d=-\sum_n \rho_{nn} \ln \rho_{nn}\nonumber
\ea
where $\rho_{nn}$ is the $n$-th diagonal element of the density matrix describing
the system. One can interpolate it to obtain a time dependent 
diagonal entropy where $\rho_{nn}(t)$ is the diagonal elements of the
density matrix in the instantaneous eigen basis. In our case, 
$\rho_{11}(t)=1-p_k(t)$ and $\rho_{22}(t)=p_k(t)$ where the excitations
for each mode $k$ are calculated in the instantaneous eigen basis.
Here, we compare the evolution of the diagonal entropy in the reverse and 
forward cases. We find that in the forward case
the diagonal entropy increases 
monotonically with time and eventually saturates to the asymptotic 
value corresponding to the von-Neuman entropy whereas in the reverse case,
a dip is observed immediately after the critical point. 

\begin{figure}
\includegraphics[height=3.2in,angle=-90]{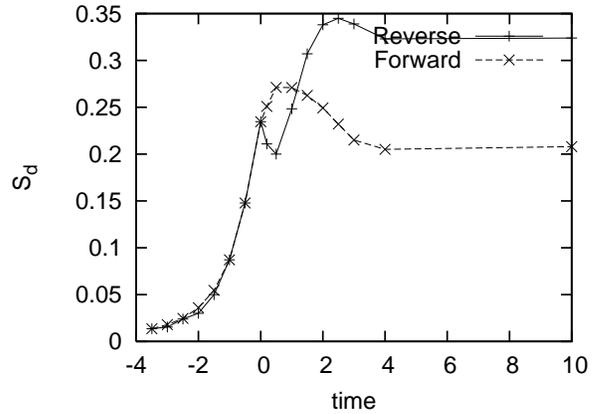}
\caption{Variation of diagonal entropy with time for $\tau=10$.}
\label{entropytime}
\end{figure}

The experimental realization of the Kitaev model has been proposed recently
in systems of ultracold atoms and molecules trapped in optical lattices
\ct{duan03}. In this proposal, each of the couplings can be independently
tuned using different microwave radiations. Once 
this is established, one can experimentally verify the reverse quenching case
by looking at the defect density which actually corresponds to the number of 
{\it bosons} in the wrong spin state.
It is also possible to investigate the spatial correlation function of 
the operator $ib_n a_{n+r}$ where $a_n$ and $b_n$ are Majorana
fermions as defined before \ct{sengupta08}. 
This spatial correlation function depends on
$p_k$ which we have already obtained for the reverse case.
Then the evolution of defect correlations can be 
detected by spatial noise correlation measurements as discussed in Ref. 46.
Also, qualitative testing of reverse quenching can be done by varying a 
magnetic field in spin gap dimer compounds such as ${{\rm BaCuSi_2 O_6}}$ which
undergo a singlet-triplet quantum phase transition at a critical field $B_c$.
Our results suggest that the density of defects in the reverse case
which correspond to residual singlets
obtained by magnetization measurement would be close to half of the forward
case.


\section{CONCLUSIONS}

In conclusion, we study the dynamics of a one-dimensional Kitaev model  using
a reverse quenching scheme, by
increasing the anisotropy parameter $J_-$ linearly 
up to its quantum critical point at $t=0$ following which $J_-$ is decreased at the same rate
to bring it back to its initial value. We provide  an exact solution of the 
Schr\"odinger equation and estimate the density of defects and the 
local entropy density in the final state. Comparison 
 of the reverse quenching results
to those of  the corresponding forward quenching case leads to a few interesting observations.
In the reverse case, the Landau-Zener transition probability $p_k$ vanishes 
for the critical mode $k=\pi/2$ at which the instantaneous energy
gap vanishes at $t=0$ whereas $p_k$ is maximum for the same mode in the
forward case \ct{sengupta08}. We show  that
$p_k$ increases linearly with $\alpha$ in small
$\alpha$ (diabatic) limit whereas our result retrieves the expected  
$1/\alpha^2$ fall of $p_k$ in the large $\alpha$ (adiabatic) limit
as predicted from quantum mechanical adiabatic theorem; $\alpha$
is the effective rate parameter as defined in the text. 
Interestingly,
the density of defects in the reverse case is close to half that of 
forward quenching.  
We have also compared the half quenching case with reverse quenching.
A close inspection of $p_k$ as a function of $k$ as shown in Fig.~6 suggests
that in the reverse case, $p_k$ gets reorganized in wave vector space with
a peak at $k=k_0\ne \pi/2$
keeping the density of defects approximately same as half quenching case.
The value of this $k_0$
shifts to $\pi/2$ for $\tau \to \infty$. We also show that the maximum value of the transition probability
is independent of $\tau$.  The local entropy density of the final state and time evolution
of the diagonal entropy density are also explored. The possibility of experimental
realization of the reverse quenching scheme using atoms trapped on optical lattices
has also been pointed out.

\begin{center}
{\bf Acknowledgement}\\
\end{center}
AD acknowledges R. Moessner and the hospitality of MPIPKS, Dresden
 where the initial
part of this work was done
The authors acknowledge Diptiman Sen and Krishnendu Sengupta 
for critical 
comments. We also thank P. P. Kurur and 
Victor Mukherjee for many fruitful 
discussions.

\appendix
\section{Details of exact calculation}
 We present here an outline of the calculational details leading to an 
exact solution for the reverse 
quenching case generalizing earlier studies \ct {landau,sei,damski06} to the
present case.
Let us first consider a general Hamiltonian in the basis $|1\ket$ and $|2\ket$
as shown below
\ba 
H=\left[ \begin{array}{cc} \epsilon_1 & \Delta \\
\Delta^* & \epsilon_2 \end{array} \right] \non 
\ea
where $\epsilon_1$ and $\epsilon_2$ are the two bare energy levels
(diagonal elements) varying  as $\sim t/2\tau$ and $-t/2\tau$ respectively
(see Eq.~(\ref{kit_mat2}) for comparison).
If $|\psi(t)\ket=c_1(t)|1\ket+c_2(t)|2\ket$, then one can write down the 
Schr\"odinger equation for $c_1(t)$ and $c_2(t)$. But before that
we redefine $c_1(t)$ and $c_2(t)$ as follows
\ba
c_1(t) &=& \tilde c_1(t) e^{-i \int_{-\infty}^{t}\epsilon_1^{'}dt'}\non\\
c_2(t) &=& \tilde c_2(t) e^{-i \int_{-\infty}^{t}\epsilon_2^{'}dt'}.
\ea
The Schr\"odinger equation for $\tilde c_2$ is
\be
i\frac{\partial}{\partial t}\tilde c_2(t)=\Delta \tilde c_1(t) e^{-i\int_{-\infty}^{t}(\epsilon_1(t')-\epsilon_2(t'))dt'}.
\ee
One more transformation of the form
\be
\tilde c_2(t) = e^{-\frac{i}{2}\int_{-\infty}^t(\epsilon_1-\epsilon_2) dt'} U_2(t)
\ee
helps us to write the equation for $\tilde c_2$ in terms of $U_2$ 
in the following form
\be
\frac{\partial^2 }{\partial t^2}U_2(t)+(\Delta^2-\frac{i}{2 \tau}+
\frac{t^2}{4 \tau^2})U_2(t)=0
\ee
\nd where we have substituted $\epsilon_1-\epsilon_2=t/\tau$. Now
redefining a new variable 
$$z=\frac{t}{\sqrt{\tau}}e^{-i\pi/4}$$
one gets,
\be
\frac{\partial^2 }{\partial z^2}U_2(z)+(m+\frac{1}{2}-\frac{z^2}{4})U_2(z)=0
\ee
\nd where $m=i\Delta^2\tau$. By all these transformations, we are able to 
recast the Schr\"odinger equation in the form of Weber Differential 
equation \ct{whittaker}
whose solutions are linear combination of well known Weber functions
$D_{-m-1}(iz)$ and $D_{-m-1}(-iz)$ $i.e.,$
\be
U_2(z)=aD_{-m-1}(iz)+bD_{-m-1}(-iz)\non\\
\ee
or going back to the notation of $\tilde c_1(t)$ and $\tilde c_2(t)$,
\ba
|\psi(t)\ket&=&\frac{i}{\Delta}[\partial_t-\frac{it}{2\tau}][aD_{-m-1}(iz)+bD_{-m-1}(-iz)]|1\ket\non\\
&+&[aD_{-m-1}(iz)+bD_{-m-1}(-iz)]|2\ket.
\label{psit}
\ea
But the initial condition demands that
at $t\to-\infty$, $|\psi(t)\ket\sim |1\ket$  forcing 
$U_2(z)$ to be a function of only $D_{-m-1}(-iz)$ as $D_{-m-1}(-iz)$ goes
to zero at $t \to -\infty$ but $D_{-m-1}(iz)$ does not as can be seen from
the following asymptotic form of Weber functions 
\ba
D_n(z) \sim e^{-\frac{1}{4}z^2}z^n-\frac{\sqrt{2\pi}}{\Gamma(-n)}e^{n\pi ir}
e^{\frac{1}4 z^2}z^{-n-1}\non\\
{\rm for}~ \frac{\pi r}{4}<arg(z)<\frac{5\pi r}{4}
\ea
\nd where $r$ is either $1$ or $-1$, and
\ba
D_n(z)\sim e^{-\frac{1}{4}z^2} z^n~~{\rm for}~ |arg(z)|<\frac{3\pi}{4}
\ea

Therefore, $a=0$ and the form of $b$ can be obtained by these asymptotic forms 
along with the initial condition,
which gives 
$b=\Delta\sqrt{\tau}e^{-\frac{\pi}{4}\Delta^2\tau}$.

Hence, Eq~(\ref{psit}), after using the derivative of Weber function,
is
\ba
&|\psi&(t\leq 0)\ket=e^{-\frac{\pi}{4}\Delta^2\tau} e^{i\frac{3\pi}{4}}
[(m+1)D_{-m-2}(-iz)\non\\
&-&izD_{-m-1}(-iz)]|1\ket +\Delta\sqrt{\tau}
e^{-\frac{\pi}{4}\Delta^2\tau} D_{-m-1}(-iz)|2\ket\non
\ea

\nd At $t=0$, the wavefunction is
\ba
|\psi(t=0)\ket&=&e^{-\frac{\pi}{4}\Delta^2 \tau}e^{i\frac{3\pi}{4}}
\frac{\sqrt{\pi} 2^{-m/2}}{\Gamma(1/2+m/2)}|1\ket\non\\
&+&\Delta\sqrt{\tau}e^{-\frac{\pi}{4}\Delta^2\tau}\sqrt{\frac{\pi}{2}}
\frac{ 2^{-m/2}}{\Gamma(1+m/2)}|2\ket
\ea
which is obtained by using the following property
$$\lim_{s\rightarrow 0}D_m(s)=2^{m/2}\frac{\sqrt{\pi}}{\Gamma(1/2-m/2)}+O(s).$$
The wave function at $t>0$, which has the effect of reversing,
should match with the wave function for $t<0$ at $t=0$. In the reverse case,
the parameters $m$ and $z$ for $t>0$ are redefined as 
\ba
m'=-i\Delta^2 \tau~{\rm and}~ z'=\frac{-it}{\sqrt{\tau}}e^{-i\pi/4}.
\ea
With these redefined $k'$ and $z'$, starting from Eq~(\ref{psit}), the 
wavefunction for $t<0$ are matched with that of $t>0$ at $t=0$ to obtain
coefficients $a$ and $b$ 
\ba
a=\frac{1}{2}\Delta\sqrt{\tau}e^{-\frac{\pi}{4}\Delta^2\tau}2^{-m}
\times \left[\frac{\Gamma(1-m/2)}{\Gamma(1+m/2)}-i\frac{\Gamma(1/2-m/2)}
{\Gamma{1/2+m/2}}\right]\non\\
b=\frac{1}{2}\Delta\sqrt{\tau}e^{-\frac{\pi}{4}\Delta^2\tau}2^{-m}
\times \left[\frac{\Gamma(1-m/2)}{\Gamma(1+m/2)}+i\frac{\Gamma(1/2-m/2)}
{\Gamma{1/2+m/2}}\right]\non
\ea
We know that at $t \to \infty$, the excited state is $|2\ket$ and hence
the coefficient $|c_2|^2$ defines the excitation probability and is equal to
\ba
c_2(t\to \infty) \sim \lim_{z\to \infty}a D_{-m'-1}(iz')+bD_{-m'-1}(-iz')
\ea
Once again, using the expression for $b$ and asymptotic expansion
of Weber function in the definition of $|c_2(t \to \infty)|^2$ 
 with properly identifying
$\tau$ and $\Delta$ for a Kitaev model, we get Eq~(\ref{eq_pk}).


\end{document}